\documentclass[12pt,thmsa]{article}
\usepackage{sw20lart}

\input tcilatex
\QQQ{Language}{
British English
}

\begin{document}

\title{Limits on Cosmological Magnetic Fields and Other Anisotropic Stresses }
\author{\and John D. Barrow \\
%EndAName
Astronomy Centre,\\
University of Sussex,\\
Brighton BN1 9QH\\
U.K.}
\maketitle
\date{}

\begin{abstract}
We discuss the cosmological evolution of matter sources with small
anisotropic pressures. This includes electric and magnetic fields,
collisionless relativistic particles, gravitons, antisymmetric axion fields
in low-energy string cosmologies, spatial curvature anisotropies, and
stresses arising from simple topological defects. In many interesting cases
the evolution displays a special critical behaviour created by the
non-linear evolution of the pressure and expansion anisotropies. The COBE
microwave sky maps are used to place strong limits of order $\Omega
_{a0}\leq 5\times 10^{-6}\Delta \ (1+z_{rec})^{-\Delta }\ $on the possible
contribution of these matter sources to the total density of the universe,
where $1\leq \Delta \leq 3$ characterises the anisotropic stress. In the
case of cosmological magnetic fields this translates into a present-day
bound of $B_0<3.4\times 10^{-9}\ (\Omega _0h_{50}^2)^{\frac 12}\mathrm{Gauss.%
}$ We explain why the limits obtained from primordial nucleosynthesis are
generally weaker than those imposed by the microwave background isotropy.
The effect of inflation on all these stresses is also calculated.
\end{abstract}

\section{Introduction}

One of the key problems of cosmology is deciding which matter fields are
present in the Universe. The most notable uncertainties govern the presence
or absence of influential matter fields which might contribute to the total
density of the Universe today, and thereby offer a resolution of the 'dark
matter'[1] problem or provide an effective cosmological constant or vacuum
stress [2] which alters the significance of the evidence governing the
process of large-scale structure formation. Other matter fields, like
magnetic fields or topological defects, can have a profound influence upon
the evolution and properties of galaxies. In the early universe the identity
of matter fields is even more uncertain. String theories are populated by
large numbers of high-mass fields, but they will not survive to influence
the late-time evolution of the Universe significantly if inflation takes
place at high energies. Inflation, of course, involves it own material
uncertainties: it requires the existence of a weakly-coupled scalar field
which evolves slowly during the early universe [3]. In all these cases the
principal problems are ones of fundamental physics -- do the postulated
matter fields exists or not?, what are the strengths of their couplings to
themselves and to other fields? Relic densities residing in isotropic or
scalar stresses are the simplest to determine. They can be calculated from a
semi-analytic solution of the Boltzmann equation if their interaction
strengths, lifetimes and masses are known. Limits on their masses,
lifetimes, and interaction cross-sections are obtained by requiring that
they do not contribute significantly more than the closure density to the
Universe today, or produce too many decay photons in a particular portion of
the electromagnetic spectrum [1], [4].

The most astrophysically interesting anisotropic stress is that of a
cosmological magnetic field. The origin of large-scale magnetic fields,
whether observed in galaxies or galaxy clusters, is still a mystery.
Intracluster fields are largely dominated by ejecta from galaxies. The
invocation of protogalactic dynamos to explain the magnitude of the galactic
field involves many uncertain assumptions but still requires a small
primordial (pregalactic) seed field. Hence the possibility of a primordial
field merits serious consideration. Other attempts to find an origin for the
field in the early universe have appealed to battery effects, the
electroweak phase transition, or to fundamental changes in the nature of the
electromagnetic interaction. All introduce further hypotheses about the
early universe or the structure of the electroweak interaction. All aim to
generate fields by causal processes when the Universe is of finite age.
Therefore, any magnetic field created by these means will exist only on very
small scales with an energy density that is a negligible fraction of the
background equilibrium radiation energy density.

Nevertheless, while such fields might still provide the seeds for non-linear
dynamos in the post-recombination era, any \textit{large-scale} magnetic
field with a strength of order $B\simeq 10^{-8}$ Gauss, comparable to that
inferred from the lowest measured intergalactic fields and close to the
observational upper limits via Faraday rotation measurements, may well be of
cosmological origin. A similar pregalactic (or protogalactic) field strength
is inferred from the detection of fields of order $10^{-6}$ Gauss in high
redshift galaxies and in damped Lyman alpha clouds, where the observed
fields are likely to have been adiabatically amplified during protogalactic
collapse. In the absence of a plausible dynamo for generating large-scale
pregalactic fields, it is of interest to reconsider the limits on a
large-scale primordial field in view of new observational constraints that
we outline below.

Primordial magnetic fields can leave observable traces of their influence on
the expansion dynamics of the Universe because they create anisotropic
pressures and these pressures require an anisotropic gravitational field to
support them. The influence of a magnetic field on reaction rates at
nucleosynthesis only limits the equivalent current epoch field to be less
than about $3\times 10^{-7}$ Gauss. We shall see that the cosmic microwave
background isotropy provides a stronger limit on the strength of a
homogeneous component of a primordial magnetic field.

We will consider the general behaviour of anisotropic stresses, of which
electric and magnetic fields are the simplest examples. We shall not assume
that inflation has taken place, although we will also calculate the effects
of inflation on their present abundances. The isotropy of the microwave
background requires any such anisotropies to be extremely small but we shall
find that their evolution is subtle because of the coupling between pressure
anisotropies and the expansion dynamics during the radiation era.
Anisotropic stresses have many possible sources --- besides cosmological
magnetic and electric fields [5], examples are provided by populations of
collisionless particles like gravitons [6], photons [7], or relativistic
neutrinos [8], [9]; long-wavelength gravitational waves [10], [11],
Yang-Mills fields [12], axion fields in low-energy string theory [13], [14];
and topological defects like cosmic strings and domain walls [15]-[17]. Our
experience with high-energy physics theories also alerts us to the
possibility that a (more) final theory of high-energy physics will contain
many other matter fields, some of which may well exert anisotropic stresses
in the universe.

Cosmological observations of the isotropy of the microwave background tell
us that the Universe is expanding almost isotropically. Therefore we can
expect to employ the high isotropy of the microwave background to constrain
the possible density of any relic anisotropic matter fields in the Universe.
We shall see that a very general analysis is possible, largely independent
of the specific identity of the physical field in question, which leads to
strong limits on the existence of homogeneous matter fields with anisotropic
pressures, irrespective of initial conditions in many cases. This analysis
exhibits a number of interesting features of the general behaviour of small
deviations from isotropy in expanding universes in the presence and absence
of a period of inflation.

In section 2 we set up the cosmological evolution equations for a universe
containing a perfect fluid and a general form of anisotropic stress. Under
the assumption that the anisotropy is small these equations can be solved.
They reveal two distinctive forms of evolution in which the abundance of the
anisotropic fluid is determined linearly or non-linearly, respectively. The
evolution can be parametrised in terms of a single parameter which
characterises the pressure anisotropy of the anisotropic fluid. In section 3
we give a number of specific examples of anisotropic fluids and explain how
this analysis also allows us to understand the evolution of the most general
anisotropic universes with 3-curvature anisotropy with respect to the
simplest examples which possess isotropic 3-curvature. In section 4 we
derive limits on the abundances of anisotropic stresses in the Universe
today by using the COBE four-year anisotropy data [18] and describe the
sampling procedure used to extract maximum information from the COBE maps.
We also consider the case of an anisotropic stress which becomes
non-relativistic after the end of the radiation era (analogous to a light
neutrino species becoming effectively massive). In section 5 we consider the
effect of a period of de Sitter or power-law inflation on the abundances of
anisotropic stresses surviving the early universe, and in section 6 we
compare the constraints we have obtained on anisotropic stresses from COBE
with those that can be derived by considering their effects on the
primordial nucleosynthesis of helium-4. The effect of the pressure
anisotropy is to slow the decay of shear anisotropies to such an extent
(logarithmic decay in time during the radiation era) that in the general
case the COBE limits are far stronger than those provided by nucleosynthesis.

\section{The Cosmological Evolution of Skew Stresses}

Consider an anisotropic universe with metric [19]

\begin{equation}
ds^2=dt^2-a^2(t)dx^2-b^2(t)dy^2-c^2(t)dz^2.  \label{eq.a}
\end{equation}
If we define the expansion rates by

\[
\alpha =\frac{\dot a}a;\text{ }\beta =\frac{\dot b}b;\text{ }\theta =\frac{%
\dot c}c, 
\]
where overdot denotes $d/dt,$ then the Einstein equations are

\begin{eqnarray}
\dot \alpha +\alpha ^2+\alpha (\beta +\theta ) &=&-8\pi G(T_1^1-\frac 12T),
\label{eq.b1} \\
\dot \beta +\beta ^2+\beta (\alpha +\theta ) &=&-8\pi G(T_2^2-\frac 12T),
\label{eq.b2} \\
\dot \theta +\theta ^2+\theta (\alpha +\beta ) &=&-8\pi G(T_3^3-\frac 12T),
\label{eq.b3} \\
\frac{\ddot a}a+\frac{\ddot b}b+\frac{\ddot c}c &=&-8\pi G(T_0^0-\frac 12T),
\label{eq.b4}
\end{eqnarray}
where $T_{a\ }^b$ is the energy-momentum tensor and henceforth we set $8\pi
G=1$.

We define the mean Hubble expansion rate, $H$, by

\begin{equation}
H=\frac{\alpha +\beta +\theta }3,  \label{eq.c}
\end{equation}
and the two relative shear anisotropy parameters by

\begin{equation}
R=\frac{\alpha -\beta }H\text{ and }S=\frac{\alpha -\theta }H\text{ .}
\label{eq.d}
\end{equation}
When $R=S=0$ the universe will be the isotropic flat Friedmann universe.

We are interested in studying the behaviour of a non-interacting combination
of a perfect fluid and a fluid with an anisotropic pressure distribution.
Thus the energy-momentum tensor for the universe is a sum of two parts:

\begin{equation}
T_b^a\equiv t_b^a+s_b^a,  \label{eq.e}
\end{equation}
where $t_b^a$ is the energy-momentum tensor of the perfect fluid, so

\begin{equation}
t_b^a=diag(\rho _{*},-p_{*},-p_{*},-p_{*})  \label{eq.f1}
\end{equation}
with equation of state 
\begin{equation}
p_{*}=(\gamma -1)\rho _{*},  \label{eq.f}
\end{equation}
and $s_b^a$ is the energy-momentum tensor of a fluid with density $\rho $
and principal pressures $p_1,p_2,$ and $p_3,$ so

\begin{equation}
s_b^a=diag(\rho ,-p_1,-p_2,-p_3).  \label{eq.g}
\end{equation}
We shall assume that the principal pressures of the anisotropic fluid are
each proportional to its density, $\rho ,\ $so that we have

\begin{equation}
(p_1,p_2,p_3)\equiv (\lambda \rho ,\nu \rho ,\mu \rho ),  \label{eq.h}
\end{equation}
where $\lambda ,\nu ,$ and $\mu $ are constants. It will be useful to define
the sum of the pressure parameters by

\begin{equation}
\Delta \equiv \lambda +\mu +\nu .  \label{eq.h1}
\end{equation}

The ansatz of (\ref{eq.h}) is clearly not the most general possible. We
could seek to extend it by allowing $\lambda ,\nu ,$ and $\mu $ to be
time-dependent quantities. However, this opens up a huge range of complex,
non-integrable behaviours. We know this because it allows the introduction
of a Yang-Mills field which produces chaotic evolution (see Barrow and Levin
[45]) equivalent to the motion of a point inside a potential bounded by four
walls formed by the rectangular hyperbolae $x^2y^2=$constant.

In the special case of an axisymmetric metric, (\ref{eq.a}), two of these
three constants would be equal. The ratio of the densities of the two fluids
is defined by 
\begin{equation}
Q\equiv \frac \rho {\rho _{*}}.  \label{eq.i}
\end{equation}
The two stresses $t_b^a$ and $s_b^a$ are assumed to be separately conserved,
so

\begin{equation}
\nabla ^bt_b^a=0=\nabla ^bs_b^a,  \label{eq.i1}
\end{equation}
the $a=0$ components of which, using (\ref{eq.f})-(\ref{eq.h}), give two
conservation equations,

\begin{eqnarray}
\dot \rho _{*}+3H\gamma \rho _{*} &=&0,  \label{eq.j1} \\
\dot \rho +\rho [\alpha (\lambda +1)+\beta (\mu +1)+\theta (\nu +1)] &=&0.
\label{eq.j2}
\end{eqnarray}
Hence, the perfect fluid density falls in proportion to the comoving volume,
as

\begin{equation}
\rho _{*}\propto \frac 1{(abc)^\gamma },  \label{eq.k}
\end{equation}
and the anisotropic fluid density falls as

\begin{equation}
\rho \propto \frac 1{a^{\lambda +1}b^{\nu +1}c^{\mu +1}}.  \label{eq.m}
\end{equation}
One can see from these two equations that $Q\propto a^{\gamma -\lambda
-1}b^{\gamma -\nu -1}c^{\gamma -\mu -1}$ admits a variety of possible
time-evolutions according to the values of the isotropic and anisotropic
pressures. The evolution equations become,

\begin{eqnarray}
\dot \alpha +\alpha ^2+\alpha (\beta +\theta ) &=&\frac 12[\rho
+p_1-p_2-p_3+(2-\gamma )\rho _{*}]\   \label{eq.n1} \\
\dot \beta +\beta ^2+\beta (\alpha +\theta ) &=&\frac 12[\rho
+p_3-p_1-p_2+(2-\gamma )\rho _{*}]\   \label{eq.n2} \\
\dot \theta +\theta ^2+\theta (\alpha +\beta ) &=&\frac 12[\rho
+p_2-p_3-p_1+(2-\gamma )\rho _{*}]\   \label{eq.n3} \\
&&\   \nonumber
\end{eqnarray}
Substituting $H,R$ and $S$ for $\alpha ,\beta ,$ and $\theta ,$ from (\ref
{eq.d}) and using (\ref{eq.n1})-(\ref{eq.n3}) and (\ref{eq.b4}), we have two
propagation equations for the anisotropies,

\begin{eqnarray}
H\dot R+R\dot H+3RH^2\ &=&p_1-p_3=(\lambda -\mu )Q\rho _{*},  \label{eq.o1}
\\
H\dot S+S\dot H+3SH^2 &=&p_1-p_2=(\lambda -\nu )Q\rho _{*},  \label{eq.o2}
\end{eqnarray}
and the conservation equations (\ref{eq.j1})-(\ref{eq.j2}) combine to give

\begin{equation}
\dot Q=\frac{qH\ }{3\ }[9(\gamma -1)-3(\lambda +\nu +\mu )+R(2\mu -\nu
-\lambda )+S(2\nu -\lambda -\mu )]  \label{eq.p}
\end{equation}
The three equations (\ref{eq.o1})-(\ref{eq.p}) completely determine the
evolution when the anisotropy is small. We are interested in the solutions
of these equations in the case where the anisotropy is realistically small.
Thus we assume that $H$ and $\rho _{*}$ take the values they would have in
an isotropic Friedmann universe containing isotropic density $\rho _{*}$
(ie. $S=R=Q=0),$ so

\begin{equation}
H=\frac 2{3\gamma t}\text{ and }\rho _{*}=\frac 4{3\gamma ^2t^2}.
\label{eq.q}
\end{equation}
This is consistent with the $\left( _0^0\right) -$ equation, (\ref{eq.b4}).
The three equations (\ref{eq.o1})-(\ref{eq.p}) then completely determine the
evolution of $R,S,$ and $Q$. They reduce to

\begin{eqnarray}
\dot R+\frac R{\gamma t}(2-\gamma ) &=&\frac{2Q(\lambda -\mu )}{\gamma t},
\label{eq.r1} \\
\dot S+\frac S{\gamma t}(2-\gamma ) &=&\frac{2Q(\lambda -\nu )}{\gamma t},
\label{eq.r2} \\
\dot Q &=&\frac{2Q\ }{9\gamma t}[9(\gamma -1)-3\Delta +R(2\mu -\nu -\lambda
)+  \label{eq.r3} \\
&&\ +S(2\nu -\lambda -\mu )].  \nonumber \\
&&  \nonumber  \label{eq.r3}
\end{eqnarray}
Axisymmetric solutions exist when $S=0$ and $\lambda =\nu .\ \ $No essential
simplification arises by imposing axial symmetry and so we shall treat the
general case with $S\neq 0.$

The system of equations (\ref{eq.r1})-(\ref{eq.r3}) gives rise to a
characteristic pattern of cosmological evolution when stresses with
anisotropic pressures are present. Equation (\ref{eq.r3}) reveals that there
is a critical condition which, if satisfied, makes the problem a
second-order stability problem; that is, if we linearised the equations
about the isotropic solution ($R=S=0)$ with zero anisotropic stress density (%
$Q=0)$ we would find a zero eigenvalue associated with the evolution of $%
Q(t) $. This critical condition is

\begin{equation}
Criticality\text{ }condition:3(\gamma -1)=\Delta .  \label{eq.s}
\end{equation}
When $3\gamma \leq 3+\Delta $ the shear distortion variables $R$ and $S$
relax towards their attractor where $\dot R=\dot S=0$ as $t\rightarrow
\infty ,$ and so we have

\begin{eqnarray}
R &=&\frac{2Q(\lambda -\mu )}{2-\gamma \ }+\delta _1t^{\frac{\gamma -2}%
\gamma }\rightarrow \frac{2Q(\lambda -\mu )}{2-\gamma \ },  \label{eq.t1} \\
S &=&\frac{2Q(\lambda -\nu )}{2-\gamma }+\delta _2t^{\frac{\gamma -2}\gamma
}\rightarrow \frac{2Q(\lambda -\nu )}{2-\gamma }.  \label{eq.t2}
\end{eqnarray}
with $\delta _1$ and $\delta _2$ constants. The $\delta _it^{\frac{\gamma -2}%
\gamma }$ terms are the contributions from the isotropic part of the
3-curvature. They fall off more rapidly than the part of the shear that is
driven by the anisotropic pressure (assuming $\lambda \neq \mu \neq \nu $).
As $t$ grows, the $\delta $ terms become negligible if $3\gamma >2\Delta $
(and recall that the isotropy is stable so long as $3\gamma \leq 3+\Delta ).$
Thus, $R$ and $S$ become proportional to $Q(t),$ which is determined by

\begin{equation}
\dot Q=\frac{2Q}{9\gamma t}\{9(\gamma -1)-3\Delta +\frac{2Q}{(2-\gamma )}\
[(\lambda -\mu )(2\mu -\nu -\lambda )+(\lambda -\nu )(2\nu -\lambda -\mu
)]\}.  \label{eq.u}
\end{equation}

When the evolution is not critical $(3(\gamma -1)\neq \Delta )$ the
evolution of $Q(t)$ is determined by the terms linear in $Q$ on the
right-hand side of (\ref{eq.u}); eqns. (\ref{eq.t1})-(\ref{eq.t2}) still
hold, but now we have

\begin{eqnarray}
Q(t) &=&Q_0t^k;\text{ }k\neq 0,  \label{eq.v1} \\
k &=&\frac 2{3\gamma }[3(\gamma -1)-\Delta ].  \label{eq.v2}
\end{eqnarray}
Our assumption, eq.(\ref{eq.q}), that the evolution of $\rho _{*}$ and $H$
follow their values in an isotropic Friedmann universe will only be
consistent at large times if $k\leq 0$, that is if

\begin{equation}
-3\leq 3(\gamma -1)\leq \Delta \leq 3.  \label{eq.w}
\end{equation}
The right-hand side of this inequality derives from the causality conditions
for signal propagation in the $i=1,2,3$ directions ($p_i\leq \rho $ so $%
\lambda \leq 1,\mu \leq 1,\nu \leq 1);$ the left-hand side arises from $%
p_i\geq -\rho ,$ which ensures the stability of the vacuum. When $k>0$ the
anisotropic stress redshifts away more slowly that the isotropic perfect
fluid on the average (over directions) and comes to dominate the expansion
dynamics, making them completely anisotropic. We do not live in such a
universe. By contrast, when $k<0,$ the isotropic stresses redshift away the
slowest and increasingly dominate the dynamics, so the gravitational effect
of the anisotropic stresses steadily diminishes. In the critical case the
average stress energy of the anisotropic stress is counter-balanced by the
isotropising effect of the perfect fluid and the shear evolution is
determined by the second-order effects of the pressure anisotropy. This
effect can be seen in the study of free neutrinos by Doroshkevich, Zeldovich
and Novikov [8] and in the study of axisymmetric magnetic fields by
Zeldovich [20].

When the evolution is critical, (\ref{eq.s}) holds, $k=0,$ and the evolution
of $Q(t)$ is decided at second-order in $Q$ by

\begin{equation}
\dot Q=\frac{4Q^2\ }{9\gamma (2-\gamma )t}\{(\lambda -\mu )(2\mu -\nu
-\lambda )+(\lambda -\nu )(2\nu -\lambda -\mu )\}.  \label{eq.x}
\end{equation}
Hence,

\begin{eqnarray}
Q(t) &=&\frac{Q_0}{1+AQ_0\ln \left( \frac t{t_1}\right) },  \label{eq.y1} \\
A &\equiv &\frac{-4\ \ }{9\gamma (2-\gamma )\ }\{(\lambda -\mu )(2\mu -\nu
-\lambda )+(\lambda -\nu )(2\nu -\lambda -\mu )\},  \label{eq.y2}
\end{eqnarray}
where $Q_0$ and $t_1$ are constants. For physically realistic stresses we
have $A>0$, and so as $t\rightarrow \infty $ the ratio of the energy
densities approaches the attractor

\begin{equation}
Q\rightarrow \frac 1{A\ln \left( \frac t{t_1}\right) },\text{ }  \label{eq.z}
\end{equation}
while the associated shear distortions approach the values given by eqns. (%
\ref{eq.t1})-(\ref{eq.t2}). Since the values of $R$ and $S$ at the epoch of
last scattering of the microwave background radiation determine the observed
temperature anisotropy we will be able to constrain the allowed value of $Q$
at the present time by placing bounds on $R$ and $S$ at last scattering and
then evolving the bounds forward to the present day.

We can also calculate the asymptotic forms for the expansion scale factors
of (\ref{eq.a}). In the critical cases, as $t\rightarrow \infty ,$ they
evolve towards

\begin{eqnarray}
a(t) &\propto &t^{\frac 2{3\gamma }}\{\ln t\}^m,  \label{eq.z1} \\
b(t) &\propto &t^{\frac 2{3\gamma }}\{\ln t\}^w,  \label{eq.z2} \\
c(t) &\propto &t^{\frac 2{3\gamma }}\{\ln t\}^n.  \label{eq.z3}
\end{eqnarray}
where the constants $m,n,w$ are defined by

\begin{equation}
m=\frac{4(2\lambda -\mu -\nu )}{9\gamma A(2-\gamma )},  \label{eq.aa1}
\end{equation}

\begin{eqnarray}
n &=&\frac{4(2\mu -\nu -\lambda )}{%
\begin{array}{c}
\begin{array}{c}
9\gamma A(2-\gamma ) \\ 
\\ 
\ 
\end{array}
\  \\ 
\ 
\end{array}
\ },  \label{eq.aa2} \\
w &=&\frac{4(2\nu -\mu -\lambda )}{%
\begin{array}{c}
9\gamma A(2-\gamma ) \\ 
\\ 
\ 
\end{array}
}.  \label{eq.aa3}
\end{eqnarray}
Some specific cases will be considered below. Note that the asymptotic form
for the scale factors, (\ref{eq.z1})-(\ref{eq.z3}) is consistent with the
principal approximation imposed by eq. (\ref{eq.q}).

In the non-critical cases (with $k<0),$ as $t\rightarrow \infty ,$ the scale
factors evolve to first order as

\begin{eqnarray}
a(t) &\propto &t^{\frac 2{3\gamma }}\{1-V_at^k\},  \label{eq.aa4} \\
b(t) &\propto &t^{\frac 2{3\gamma }}\{1-V_bt^k\},  \label{eq.aa5} \\
c(t) &\propto &t^{\frac 2{3\gamma }}\{1-V_ct^k\},  \label{eq.aa6} \\
&&  \nonumber
\end{eqnarray}
where the constants $V_a,V_b,$ and $V_c$ are defined by,

\begin{equation}
\{V_a,V_b,V_c\}\equiv \frac{-4q_0}{9k\gamma (2-\gamma )}\times \{\mu +\nu
-2\lambda ,\lambda +\mu -2\nu ,\lambda +\nu -2\mu \}.  \label{eq.ab1}
\end{equation}

These asymptotes reveal the different behaviour in the critical cases which
produces the logarithmic decay of the shear. In the non-critical cases the
anisotropic perturbations to the scale factors decay as power laws in time
since $k<0.$ 
\[
\]

\section{Some Particular Skew Stresses}

There are many examples of anisotropic stresses to which the analysis of the
last section might be applied. We shall consider some of the most
interesting. In each case we can determine the characteristics of the
'critical' state in which the evolution of the anisotropy is determined by
the non-linear coupling with the anisotropic stress. In section 4 we will go
on to calculate the observed microwave background temperature anisotropy.
The most familiar example of an anisotropic stress is that of an
electromagnetic field.

\subsection{Magnetic or Electric fields}

In this case the anisotropic energy-momentum tensor $s_a^b$ of eqn. (\ref
{eq.g}) has a simple form [20], [22], [24]. For example, for a pure magnetic
field of strength $B$ directed along the z-axis, we have $%
T_0^0=T_3^3=-T_1^1=-T_2^2=B^2/8\pi $ and so this corresponds to the choice 
\begin{equation}
\lambda =\nu =-\mu =1  \label{eq.A}
\end{equation}
Hence,

\begin{equation}
\Delta (magnetic)=\Delta (electric)=1  \label{eq.A1}
\end{equation}

Therefore, from eqn.(\ref{eq.s}), we see that the criticality condition is
obeyed when the background universe is radiation dominated. Thus, in the
standard model of the early universe, the evolution of magnetic (or
electric) fields will exhibit the non-linear logarithmic decay found in
equations (\ref{eq.x})-(\ref{eq.z}). This was first pointed out by Zeldovich
[20], and can be identified in the calculations of Shikin [21] and Collins
[22]. Barrow, Ferreira, and Silk [23] used the COBE data set [18] to place
constraints on the allowed strength of any cosmological magnetic field.

As a specific example of a critical case, a radiation dominated universe ($%
\gamma =4/3)$ containing a magnetic field aligned along the z-axis, $\lambda
=\nu =-\mu =1,$ gives $A=4$ in (\ref{eq.y2}) and hence

\begin{eqnarray}
a(t) &\propto &b(t)\propto t^{\frac 12}\{\ln t\}^{\frac 14}  \label{eq.ab3}
\\
c(t) &\propto &t^{\frac 12}\{\ln t\}^{-\frac 12}  \label{eq.ab4}
\end{eqnarray}
Notice that the volume expansion goes as in the isotropic radiation
universe, $abc\propto t^{\frac 32\text{ }}$ in accord with the principal
approximation stipulated by (\ref{eq.q}).

As an example of a non-critical case consider a pure magnetic field aligned
along the $z$-axis of a dust universe ($\gamma =1)$. We have $k=-2/3,$ and
so at late times

\begin{eqnarray*}
a(t) &\propto &b(t)\propto t^{\frac 23}\{1-\frac{4Q_0}{3t^{\frac 23}}+...\},
\\
c(t) &\propto &t^{\frac 23}\{1+\frac{8Q_0}{3t^{\frac 23}}+...\}.
\end{eqnarray*}
This behaviour can be seen in the exact magnetic dust solutions of Thorne
[24] and Doroshkevich [25]. Other studies of the evolution of cosmological
magnetic fields in anisotropic universes can be found in refs. [28].

The axion field in low-energy string theory [13], [14] also creates stresses
of this characteristic form and a source-free magnetic field has been used
to study the possibility of dimensional reduction in cosmologies with
additional spatial dimensions near the Planck time by Linde and Zelnikov
[26] and Yearsley and Barrow [27].

\subsection{General Trace-free stresses}

The magnetic field case is just one of a whole class of skew fields which
exhibit non-linear evolution during the radiation era. Any anisotropic
stress which has a trace-free energy-momentum tensor, $s_a^b$, will have

\begin{equation}
\Delta =1  \label{eq.B}
\end{equation}
and will exhibit critical evolution in the presence of isotropic black body
radiation with $\gamma =4/3.$ This case includes free-streaming gravitons
produced at $t_{p\ell }\sim 10^{-43}s$ which are collisionless at $%
t>10t_{p\ell }$ because of the weakness of the gravitational interaction
mediating graviton scatterings [6]. It is also likely to include all
asymptotically-free particles at energies exceeding $\sim 10^{15}GeV$ when
interparticle scatterings, decays and inverse decays have interaction rates
slower than the expansion rate of the universe, $H$. In all cases defined by 
$\Delta =1$ the shear to Hubble expansion rate decays only logarithmically
during the radiation era,

\begin{equation}
R\propto S\propto \frac 1{\ln \left( \frac t{t_1}\right) }.  \label{eq.B2}
\end{equation}

In the dust era the critical condition will not continue to be met for
trace-free fields. Although the evolution of $Q,R,$ and $S$ is now
determined at linear order in (\ref{eq.t1}), (\ref{eq.t2}), and (\ref{eq.z}%
), there is still a significant slowing of the decay of isotropy by the
anisotropic stresses compared to the case where anisotropic stresses are
absent. We see that when $\gamma =1$ we have

\begin{equation}
\frac \rho {\rho _{*}}\propto R\propto S\propto \frac 1{t^{2/3}}
\label{eq.C}
\end{equation}
compared to a decay of $R\propto S\propto t^{-1}$ when anisotropic stresses
are absent $(Q=0)$ or ignored$.$ The study by Maartens \textit{et al }[29]
of the evolution of anisotropies in a dust-dominated universe containing
possible anisotropic stresses at second order (described by a tensor $\pi
_a^b$ in reference [28] which is equivalent to the $s_a^b$ used here)
displays this same slowing of the shear decay (the discussion of ref. [41]
omits this consideration).

In general, we can see that if the criticality condition is satisfied in the
radiation era, when $\gamma =4/3$, then it cannot be satisfied during the
dust era that follows, when $\gamma =1.$ However, an interesting situation
can arise if the source of the trace-free stress is a population of
particles which are relativistic above some energy (so $\Delta =1$ there)
and non-relativistic when the universe cools below this energy scale (so $%
\Delta =0$ there). Thus there can be a change in the value of $\Delta $ with
time. We shall examine this case in section 4.

Some matter fields with anisotropic pressures, like Yang-Mills fields [12],
correspond to an anisotropic fluid with time-dependent $\lambda ,\mu ,$ and $%
\nu $. But if the evolution is slow enough, it is well-approximated by the
model with constant values of $\lambda ,\mu ,$ and $\nu $ used here.

\subsection{Long-wavelength gravitational waves}

In the last section we considered the evolution of anisotropic stresses in
the simplest flat ($\Omega _0=1$) anisotropic cosmological model of Bianchi
type I with the metric (\ref{eq.a}). The most general anisotropic universes
of this curvature are of Bianchi type $VII_0$ and they have an Einstein
tensor that can be decomposed into a sum of two pieces: one corresponding to
the Einstein tensor for the simple type I geometry, the other to a piece
that describes additional long-wavelength gravitational waves [10], [11],
[30]. These waves create anisotropies in the 3-curvature of the universe in
addition to the simple expansion-rate anisotropies present in the Bianchi I
universe (which has isotropic 3-curvature). However, the contribution by the
long-wavelength gravitational waves can be moved to the other side of the
Einstein equations and reinterpreted as an additional 'effective'
energy-momentum tensor describing a 'fluid' of gravitational waves. It has
vanishing trace. This decomposition means that any general flat (or open)
Bianchi type universe of type $VII_0$ (or $VII_h$) containing an isotropic
perfect fluid, (\ref{eq.f}) behaves like a Bianchi type I (or type V)
universe containing that fluid plus an additional traceless anisotropic
fluid. The parameters $\lambda ,\mu ,$ and $\nu $ are approximately constant
when the anisotropies are small. Hence the general evolution of anisotropic
universes with anisotropic curvature containing isotropic radiation will
exhibit the same characteristic logarithmic decay of the shear anisotropy
given by eqns. (\ref{eq.t1}), (\ref{eq.t2}), and (\ref{eq.z}) during the
radiation era. This behaviour appears in\textit{\ }Collins and Hawking [31],
Doroshkevich \textit{et al} [32], and other authors, [33], [34]. During the
dust era any anisotropic curvature modes will behave like trace-free
stresses and, although the evolution will no longer be critical, the shear
anisotropy will fall more slowly ($R\propto S\propto t^{-2/3}$ ) than in
simple isotropic universes with isotropic curvature and no anisotropic
stresses (where $R\propto S\propto t^{1-2\gamma }$).

\subsection{Strings and Walls}

Topological defects provide another class of anisotropic energy-momentum
tensors which fall into the category of stress modelled by (\ref{eq.h}) for
part of their evolution. The energy-momentum tensors of string sources were
also considered more generally by Stachel [16], Marder and Israel [17]. The
specific description of line stresses created by topological defects is
reviewed in ref. [15]. An infinite string with mass per unit length $\mu $
extending in the x-direction contributes an anisotropic stress-tensor $%
s_a^b=\mu \delta (z)\delta (y)diag(1,1,0,0)$; that is$\ $

\begin{equation}
\lambda =-1,\text{ }\mu =\nu =0\rightarrow \Delta (string)=-1.  \label{eq.D}
\end{equation}

Their evolution could therefore only be critical in a universe containing a
perfect fluid with equation of state $p_{*}=-\rho _{*}/3.$ It is worth
noting that this corresponds to the evolution of the curvature term in the
Friedmann equation when the universe is open. Thus we would expect the
string stress to evolve critically during the late curvature-dominated stage
($1+z<\Omega _0^{-1}$) of an open universe, during a curvature-dominated
pre-inflationary phase, or during any period when quantum matter fields with 
$\gamma =1/3$ dominate the expansion.

For slow-moving infinite planar domain walls of constant surface density in
the $x-y$ plane $,$ the stress corresponds $s_a^b\propto \eta
(z)diag(1,1,1,0),$ where $\eta (z)$ is a local bell-shaped curve centered on 
$z=0$ [15]$,$and corresponds to the choice

\begin{equation}
\lambda =\nu =-1,\text{ }\mu =0\rightarrow \Delta (wall)=-2.  \label{eq.E}
\end{equation}
The wall stress would only be critical if the equation of state of the
background universe is $p_{*}=-2\rho _{*}/3.$

These descriptions do not include the complicated effects of the non-linear
evolution of a population of open strings and loops which must include
intersections, kinks, gravitational collapse and gravitational radiation.

\subsection{General classification and energy conditions}

When evaluating the form of the residual anisotropy and energy density
created by anisotropic cosmological stresses it is most convenient to
classify stresses by the value of $\Delta .$ We can circumscribe the likely
range of realistic $\Delta $ values by considering some general restrictions
on the energy-momentum tensor. The dominant energy conditions [35] require
us to impose the physical limits

\begin{equation}
\left| \lambda \right| \leq 1,\text{ }\left| \mu \right| \leq 1,\text{ }%
\left| \nu \right| \leq 1.  \label{eq.H}
\end{equation}
Hence we have the bounds

\begin{equation}
-3\leq \Delta \leq 3.  \label{eq.I}
\end{equation}

If the strong-energy condition [35] were imposed on the energy-momentum
tensor $s_a^b$ then we would have a stronger restriction

\begin{equation}
\Delta \geq -1.  \label{eq.J}
\end{equation}
This condition is violated by string and wall stresses and necessarily by
any isotropic stress which drives inflation $(0\leq \gamma <2/3)$ since $%
\Delta (wall)=-2$ and $\Delta (string)=-1.$ We see from (\ref{eq.v1})-(\ref
{eq.v2}) that there can only be approach to isotropy at late times in the
radiation era if $\Delta \geq 1.$ When $0<\Delta <1$ the expansion
approaches isotropy during the dust era but not during the radiation era.
When $\Delta \leq 0$ isotropy is unstable during both the dust and radiation
eras.

\section{Microwave background limits}

Let us consider the simplest realistic case in which the Universe is
radiation dominated until $1+z_{eq}=2.4\times 10^4\Omega _0h_0^2$ and then
dust dominated thereafter; here; $\Omega _0\leq 1$ is the cosmological
density parameter and $h_0$ is the present value of the Hubble constant in
units of $100Kms^{-1}Mpc^{-1}.$ We shall assume that the microwave
background was last scattered at a redshift $z_{rec}$, where, in the absence
of reheating of the cosmic medium,

\begin{equation}
1+z_{rec}=1100.  \label{eq.K}
\end{equation}
If there is reionization of the universe then last scattering can be delayed
until $1+z_{rec}=39(\Omega _{b0}h_0)^{-1}$ for $\Omega _0z_{rec}<<1,$ where $%
\Omega _{b0}$ is the present baryon density parameter [36].

$_{}$The microwave background temperature anisotropy is determined by the
values of shear distortions $R$ and $S$ at the redshift $z_{rec}.$ The
evolution of the photon temperature in the $x,y,$ and $z$ directions is
given by

\begin{eqnarray}
T_x &=&T_0\frac{a_0}{a(t)}=T_0\exp \{-\int \alpha dt\},  \label{eq.L1} \\
T_y &=&T_0\frac{b_0}{b(t)}=T_0\exp \{-\int \beta dt\},  \label{eq.L2} \\
T_z &=&T_0\frac{c_0}{c(t)}=T_0\exp \{-\int \theta dt\}.  \label{eq.L3}
\end{eqnarray}
If we define the temperature anisotropy by

\begin{equation}
\frac{\delta T}T\equiv \frac{(T_x^{}-T_y)\ +(T_x-T_z)^{}}{T_0}  \label{eq.M}
\end{equation}
then, for small anisotropies (so $\exp \{-\int \alpha dt\}\simeq 1-\int
\alpha dt\}$ etc), using the definitions of (\ref{eq.d}) in (\ref{eq.M}), we
have

\begin{equation}
\frac{\delta T}T=-H\int (R+S)dt\ ^{\ }  \label{eq.O}
\end{equation}
Since recombination always occurs in the dust era, $H=2/3t,$and the observed
microwave background anisotropy will be

\begin{equation}
\frac{\delta T}T=\ \left[ \frac{R+S}\Delta \right] _{rec}\times f(\vartheta
,\phi ,\Omega _0).  \label{eq.P}
\end{equation}
where $f$ $\sim O(1)$ is a pattern factor taking into account non-gaussian
statistical factors and the possible pattern structure created by
complicated forms of anisotropy in the general case [13], [23],[37].
Typically, in the most general homogeneous flat universes the pattern
combines a distorted quadrupole with a spiral geodesic motion. In ref. [23]
a detailed discussion of the sampling statistics of the microwave background
temperature distribution and the non-gaussian nature of the anisotropic
pattern was given. This enables limits to be derived from the whole COBE sky
map rather than, say, simply from the quadrupole as is generally done.

The problem of constraining global anisotropy is substantially different
from the traditional statistical task of estimating parameters in Gaussian
models. In the latter case, the ensemble is entirely characterized by the
power spectrum while in the former, a given set of parameters corresponds to
a completely specified pattern in the sky, up to an arbitrary rotation. This
problem was dealt with in some detail in Bunn et al [33]. A brief outline of
our procedure is as follows. One can model the microwave background signal
as the sum of two components: a \textit{statistically isotropic} Gaussian
random field $\Delta T_I$, which we assumed to have a scale invariant power
spectrum on the scales we are interested in, and a \textit{global}, \textit{%
anisotropic} pattern, $\Delta T_A$, which is uniquely defined by the set of
parameters $x$ (which measures the spiral 3-curvature anisotropy), $\Omega
_0,\,h_{50}$, $(\sigma /H)_0,$ and $\theta $, $\phi $ (its orientation on
the sky). Each pixel of a data set of measured microwave background
anisotropies is given by $d_i=(\Delta T\star \beta )(\mathbf{r}_i)+N_i$
where $\beta $ represents the DMR beam pattern, $\mathbf{r}_i$ is a unit
vector pointing in the direction of pixel $i$, $N_i$ is the noise in pixel $%
i $ and $\star $ is the convolution operator. To an excellent approximation,
one can assume that $N$ is Gaussian ``white'' noise, i.e. $\langle
N_iN_j\rangle =s_i^2\delta _{ij}$.

Our task is, given a pair ($x$, $\Omega _0$), to find the orientation ($%
\theta $, $\phi $) which allows the maximum observed value of $(\sigma /H)_0$%
. One can do this using standard frequentist statistical methods: we define
a goodness-of-fit statistic that depends on the data, compute its value for
the actual data, and then compute the probability that a random data set
would have given a value as good as the actual data. In Bunn et al [33], $%
\eta $ was defined to be: 
\begin{eqnarray}
&&  \nonumber  \label{eq.IN1} \\
\eta &=&\min_{\sigma ,\theta ,\phi }\eta _1;\text{ where }\eta _1={\frac{%
\Delta _0^2-\Delta _1^2}{\Delta _0^2};}  \label{eq.in1}
\end{eqnarray}
$\Delta _0^2$ is the noise-weighted mean-square value of the data and $%
\Delta _1^2$ is the noise-weighted mean-square value of the residuals after
we have subtracted off the anisotropic part. Note that removing the
incorrect anisotropic portion will only increase the residuals so the
difference between the two terms is an obvious choice for a goodness-of-fit.
Dividing by $\Delta _0^2$ ensures a weak dependence on the amplitude of the
isotropic component, while defining the statistic as the minimum of $\eta _1$
allows us to deal with the uncertainty in ($\theta ,\phi $).

This statistical method was applied to the 4-yr COBE DMR data-set. The two
53 GHz and the two 90 GHz maps were averaged together, each pixel weighted
by the inverse square of the noise level, to reduce the noise level in the
average map. All pixels within the Galactic cut were removed so as to reduce
Galactic contamination, and a best-fit monopole and dipole were subtracted
out. The map was degraded from pixelization 6 to pixelization 5 (i.e.
binning pixels in groups of four). Simulations were performed for a set of
models from the ($\Omega _0,x$) plane; for each choice of the three
parameters ($\Omega _0$, $x$ and $(\sigma /H)_0$) approximately 200 to 500
random DMR sets were generated, so allowing us to determine an approximate
fit to the probability distribution function of $\eta $.

This leads to bounds on $f$ in flat and open universes of

\begin{equation}
0.6<f<2.2.  \label{eq.P1}
\end{equation}

This bound can be improved by a factor of $\sqrt{3}$ if one considers the
results from Kogut et al [42]. In this case, a slightly different
goodness-of-fit statistic is used: instead of working with the
noise-weighted quantities, $\Delta _0^2$ and $\Delta _1^2$, the authors
chose to weight the pixels with the covariance matrix of the total Gaussian
components (i.e. the noise and isotropic cosmological components).

The discussion above shows that the $\Delta \geq 1$ case is the realistic
one which allows evolution towards isotropy at late times. The observed
anisotropy is therefore given in terms of the present value of the density
ratio, $Q(t_0)$, by

\begin{equation}
\frac{\delta T}T=\frac{2Q(t_0)f(2\lambda -\mu -\nu )(1+z_{rec})^\Delta }%
\Delta  \label{eq.Q}
\end{equation}
$\ $ If we take the COBE 4-year data set to impose a limit of $\delta
T/T\leq 10^{-5}$ on contributions by the anisotropic fluid, and use (\ref
{eq.P1}), then the present density parameter of the anisotropic fluid, $%
\Omega _{a0},$ is limited by

\begin{equation}
\Omega _{a0}\leq \frac{8.3\times 10^{-6}\Delta \ }{(2\lambda -\mu -\nu
)(1+z_{rec})^\Delta }=\left( \frac \Delta {2\lambda -\mu -\nu }\right)
\left( \ \frac{1100}{1+z_{rec}}\right) ^\Delta \times 10^{-5.08-3.04\Delta }
\label{eq.R}
\end{equation}

Since $1\leq \Delta \leq 3$ for realistic fluids we can examine the extremes
of this limit which is strongest when there is no reionization of the
Universe at $z<<1100$ because reionization allows a longer period of
power-law decay of $R,S,$ and $Q$, hence a smaller residual effect on the
microwave background isotropy. In the two extreme cases we have limits of

\begin{eqnarray}
\Delta &=&1:\Omega _{a0}\leq 3.7\times 10^{-9}\times \ \left( \frac{1100}{%
1+z_{rec}}\right)  \label{eq.S1} \\
&&  \nonumber  \label{eq.S2} \\
\Delta &=&3:\Omega _{a0}\leq 9.5\times 10^{-15}\times \ \left( \frac{1100}{%
1+z_{rec}}\right) ^3  \label{eq.S2}
\end{eqnarray}
The $\Delta =1$ case with the choice of (\ref{eq.A}) corresponds to limits
on the present cosmological energy density allowed in magnetic (or electric
fields) which was studied in [20] and in [23] where the most general
evolution of anisotropy was included. Note that these limits are far
stronger than those that are generally obtained for isotropic forms of dark
matter by imposition of astronomical limits on the maximum total density,
the age of the Universe, or the deceleration parameter [1], [4].

The most conservative limit on the cosmological magnetic field arises when
we assume that the whole anisotropy is contributed by the magnetic field
stresses. This gives us a final bound on the magnitude of the magnetic field
today of This gives a limit of

\begin{equation}
B_0<2.3\times 10^{-9}f^{\frac 12}\ (\Omega _0h_{50}^2)^{\frac 12}\mathrm{%
Gauss}  \label{eq.S3}
\end{equation}

Adams et al [43] have used the weak nucleosynthesis limit of Grasso and
Rubinstein of $B_0\leq 3\times 10^{-7}$ Gauss obtained for random fields to
argue that a cosmological magnetic field could lead to observable
distortions of the acoustic peaks in the microwave background. Our limit on $%
B_0$ rules out any observable effect of a homogeneous magnetic field on the
acoustic peaks. A large scale, inhomogeneous, magnetic field may, however,
introduce observable distortions in the acoustic peaks. Our limit permits a
field strength of $10^{-9}$ Gauss required to induce a measurable Faraday
rotation in the polarization of the microwave background [44].

The limit (\ref{eq.S3}) is not strong enough to exclude a cosmological
origin for galactic magnetic fields from purely adiabatic compression of a
background field down to galactic scales with no significant enhancement
from a dynamo. If that were the source of galactic fields then one would
expect significant fields to exist in elliptical galaxies as well as
spirals. If galactic fields were amplified significantly by dynamo action
then one would not expect significant ordered fields to exist over large
scales within ellipticals, although they could exist in local disordered
forms following ejection from stars.

\subsection{Evolution of anisotropic dark matter with a characteristic
energy scale}

We are familiar with the standard picture of the cosmological evolution of
light ($m<<1MeV)$ weakly interacting particles [1], [4]. When $T>m$ they
behave like massless particles with their number density similar to that in
photons, up to statistical weight factors. Their energy density redshifts
away like $(1+z)^4$ until the temperature falls to the value of their rest
mass. Thereafter they behave like non-relativistic massive particles and
their energy density redshifts away more slowly, as $(1+z)^3.$ We can
consider an analogous scenario for anisotropic stresses. Suppose we have an
anisotropic fluid which behaves relativistically until the temperature falls
to some value $T_{+}<10^4K\approx 1eV$ and then behaves non-relativistically
at lower temperatures. This situation corresponds to following the evolution
with $\Delta =1$ for $T\geq T_{+}$ and then with $\Delta =0$ for $T<T_{+}.$
The evolution is complicated by the fact that it will be critical during the
radiation era but non-critical during the first stage of the dust era when
the temperature exceeds $T_{+}$ when $Q$ will decay in accord with (\ref
{eq.v2}), before becoming critical again during the remainder of the dust
era until the present when $Q$ will decay logarithmically in accord with (%
\ref{eq.z}). In this case the final density of the anisotropic dark matter
is constrained by the microwave background isotropy to have

\begin{equation}
\Omega _{a0}\leq \ 1.2f^{-1}\times 10^{-4}\left( \frac{1100}{1+z_{rec}}%
\right) \times \frac 1{A\ln (1+z_{+})}  \label{eq.T1}
\end{equation}
$\ $ where

\begin{equation}
A\equiv \frac 49\{(\mu -\lambda )(2\mu -\nu -\lambda )+(\nu -\lambda )(2\nu
-\lambda -\mu )\}  \label{eq.T2}
\end{equation}
For example, if the anisotropic matter has a characteristic energy scale $m$
then, since $m=T_{+}=2.4\times 10^{-4}(1+z_{+})$ $eV,$ we have (picking $%
\lambda =\nu =-\mu =1$ so $A=32/9$), with (\ref{eq.P1}), that,

\begin{equation}
\Omega _{a0}\leq \ 5.7\times 10^{-5}\left( \frac{1100}{1+z_{rec}}\right)
\times \frac 1{\ (8.3+\ln M_{eV})}  \label{eq.T3}
\end{equation}
and there is a very weak dependence on the mass scale $m_{eV}\equiv (m/1eV).$
For $m>1eV$ we have, roughly, that

\begin{equation}
\Omega _{a0}\leq \ 6.9\times 10^{-6}\left( \frac{1100}{1+z_{rec}}\right)
\label{eq.T4}
\end{equation}

Therefore these fields are always constrained by the microwave background
anisotropy to be a negligible contributor to the total density of the
Universe.

\section{The Effects of Inflation}

So far we have ignored the consequences of any period of inflation in the
very early universe [3]. The equations (\ref{eq.r1})-(\ref{eq.t1}) hold for
the case of generalised (power-law) inflation with $0<\gamma <2/3$, [37].
However, for de Sitter inflation with $\gamma =0,$ the isotropic density
drives the inflation with $\rho _{*}=3H_0^2=$ constant, and they are changed
to

\begin{eqnarray}
\dot R+3RH_0 &=&3H_0Q(\lambda -\mu ),  \label{eq.U1} \\
\dot S+3SH_0 &=&3H_0Q(\lambda -\nu ),  \label{eq.U2} \\
\dot Q &=&\frac{\ QH_0\ }3[-9-3\Delta +R(2\mu -\nu -\lambda )+S(2\nu
-\lambda -\mu )].  \label{eq.U3}
\end{eqnarray}

This system can only be critical for the evolution of $Q$ only if $\Delta
=-3 $ but this cannot occur for an anisotropic fluid subject to (\ref{eq.H}%
). As $t\rightarrow \infty $ we have

\begin{eqnarray}
R &\rightarrow &q(\lambda -\mu )+\delta _1\exp (-3H_0t)  \label{eq.V1} \\
S &\rightarrow &q(\lambda -\nu )+\delta _2\exp (-3H_0t)  \label{eq.V2} \\
Q &\rightarrow &Q_0\exp [-(3+\Delta )H_0t]  \label{eq.V3}
\end{eqnarray}
Thus, since $\Delta +3>0$ the contribution made to the shear by the
anisotropic pressure decays. Moreover, we see that the contribution by the
isotropic curvature $(\delta )$ terms to the shear dominates the
contribution by the pressure anisotropy at late times if $\Delta >0.$ In
both cases the shear anisotropy decays away exponentially fast. This is in
accord with the expectations of the cosmic no hair theorems [38] because the
anisotropic stresses considered here obey the strong energy condition when $%
\Delta >-3.$ Thus if $N$ e-folds of de Sitter inflation occur in the very
early universe, then the values of the shear distortion parameters, $R$ and $%
S,$ are each depleted by a factor $\exp (-3N)$ whilst the relative density
in the anisotropic fluid, $q$, is reduced by a factor $\exp \{-(3+\Delta
)N\} $.

Similarly, in the case of power-law inflation ($0<\gamma <2/3$), the
isotropic curvature mode will dominate the late time evolution of the shear
during the inflationary phase if

\begin{eqnarray}
\gamma &<&\frac{2\Delta }3.  \label{eq.V4} \\
&&  \nonumber
\end{eqnarray}

If anisotropic stresses can be generated at the end of inflation then the
analysis of the previous sections applies.

\section{Primordial Nucleosynthesis}

It is instructive to consider the question of whether primordial
nucleosynthesis can provide stronger limits than microwave background
isotropy on the densities of anisotropic stresses in the Universe. The key
issue that decides this is whether the microwave background constraint on
the expansion anisotropy created by an anisotropic energy density, which is
of order $10^{-5}$ and imposed at a redshift $z_{rec}\sim 10^3$ or lower (in
the event of reionization), is stronger than a limit of order $1-10^{-1}$on
changes to the mean expansion rate imposed at the epoch of neutron-proton
freeze-out, $z_{fr}\sim 10^{10}.$ The issue is decided by knowing how fast
the anisotropic stresses decay with time between $z_{fr}$ and $z_{rec}$.

In the simplest anisotropic universes, which have isotropic 3-curvature, the
shear anisotropy falls like the that of the $\delta $ mode in eqns. (\ref
{eq.t1})-(\ref{eq.t2}). Consider first the simple isotropic curvature case
with no anisotropic stress, so $Q=0$. The evolution of the shear to Hubble
rate parameters, $R$ and $S,$ during the dust and radiation era of a
universe with $\Omega _0=1$ is

\begin{eqnarray}
z &>&z_{eq}:R\propto S\propto t^{-\frac 12}\propto 1+z  \label{eq.W1} \\
z &<&z_{eq}:R\propto S\propto t^{-1}\propto (1+z)^{\frac 32}  \label{eq.W2}
\end{eqnarray}

Hence, if nucleosynthesis gives an upper limit of $\sim 0.2$ (roughly
equivalent to adding one neutrino type to the standard three-neutrino model)
on the values of $\left| R\right| $ and $\left| S\right| $ at the redshift
of neutron-proton freeze-out, $z_{fr},\ $this corresponds to an upper limit
at the time of last scattering of $0.2(1+z_{fr})^{-1}(1+z_{eq})^{-\frac 1{\
2}}(1+z_{rec})^{\frac 32},$ so we have

\begin{equation}
R_{rec}\sim S_{rec}<4.6\times 10^{-9}(\Omega _0h_0^2)^{\frac 12}\times \
\left( \frac{1+z_{rec}}{1100}\right) ^{\frac 32}  \label{eq.W3}
\end{equation}
Since the microwave background gives a limit of $\delta T/T\sim
(R+S)_{rec}\leq 10^{-5},$[18], we see that the nucleosynthesis limit is much
stronger when the 3-curvature is isotropic, as first pointed out by Barrow
[39]; nor are conceivable improvements in microwave receiver sensitivity
ever likely to close to gap of $10^4$ that exists between the
nucleosynthesis and microwave limits on the isotropic 3-curvature $\delta $
modes. These effects of these simple anisotropy modes on nucleosynthesis
were considered in the papers of Hawking and Tayler [40] and Thorne [24].

However, the situation changes when we consider the most general forms of
anisotropy with anisotropic curvature or when matter is present with
anisotropic pressures. If the evolution will be critical during the
radiation era the anisotropy falls off so slowly during the period before $%
z_{eq}$ that the microwave background limits become stronger than the
nucleosynthesis limits. This effect was considered in both the dust and
radiation eras by Barrow [39] in the evolution of Bianchi type $VII$
universes close to isotropy but the observational limits were 100 times
weaker then. For example, in the interesting case where $\Delta =1$ the
evolution of anisotropies follows the form, (\ref{eq.B2})-(\ref{eq.C}),

\begin{eqnarray}
z &>&z_{eq}:R\propto S\propto \ \frac 1{\ln (\ \frac{z_{fr}}z)},
\label{eq.X1} \\
z &<&z_{eq}:R\propto S\propto 1+z.  \label{eq.X2}
\end{eqnarray}
Hence, the nucleosynthesis limit translates into an upper limit on $R$ and $%
S $ at the redshift of last scattering of only

\begin{equation}
R_{rec}\sim S_{rec}<7.1\times 10^{-4}(\Omega _0h_0^2)^{-1}\times \ \left( 
\frac{1+z_{rec}}{1100}\right) .  \label{eq.X3}
\end{equation}
We see that this is never stronger than the microwave background limit of $%
\leq 10^{-5}$ for any possible redshift of last scattering. Alternatively,
we might restate this result as follows: it is possible for anisotropic
fluids to create a measurable temperature anisotropy in the microwave
background radiation without having any significant effect upon the
primordial nucleosynthesis of helium-4.

\section{Discussion}

We have described the cosmological evolution of a very general class of
anisotropic stresses in the presence of an isotropic perfect fluid. Such
anisotropic stresses encompass a wide range of physically interesting cases,
including those of cosmological electric and magnetic fields, a variety of
topological defects, gravitons, and other populations of collisionless
particles. When the universe is close to isotropy the mean expansion rate
behaves as in the isotropic Friedmann universe to leading order, but the
density of the anisotropic stresses is coupled to the expansion anisotropy
in an interesting way. It was found that there are two possible forms for
the evolution. In the \textit{critical} case the density of anisotropic
matter is determined by its non-linear coupling to the expansion anisotropy
and density of anisotropic stress falls only logarithmically relative to the
isotropic background density. By contrast, in the \textit{non-critical} case
the density falls as a power-law in time relative to the background density.
In all cases the microwave background anisotropy can be used to place a
limit on the density of matter that could be residing in the universe today
in forms with anisotropic pressures. These limits arise because their
evolution approaches an asymptotic attractor in which the temperature
anisotropy produced in the microwave background by the anisotropic stresses
is simply related to their density relative to the isotropic background
density. The limits obtained on the possible cosmological density of matter
of this form are far stronger than conventional limits on isotropic forms of
dark matter because they make use of the microwave isotropy limits $(\delta
T/T\leq 10^{-5})$ rather than the far weaker limits from observations of the
Hubble flow and age of the Universe $(\Omega _0<O(1)).$ We went on to
explain why the limits that can be impose upon anisotropic stresses by the
microwave background isotropy measurements are in general stronger than
those arising from the limits on the primordial nucleosynthesis of helium-4.

\textbf{Acknowledgements}. The author is supported by PPARC. I would like to
thank Pedro Ferreira and Joe Silk for discussions.

\textbf{References}

[1] G. Jungman, M. Kamionkowski and K.Griest, Phys. Reports \textbf{267},
195 (1996).

[2] T. Padmanabhan, \textit{Structure Formation in the Universe}, Cambridge
UP, Cambridge, 1993.

[3] A. Guth, Phys. Rev. D\textit{\ }\textbf{23, }347 (1981).

[4] E. Kolb and M.S. Turner, \textit{The Early Universe}, Addison Wesley:
Redwood City (1990).

[5] Y. B. Zeldovich, Sov. Phys. JETP \textbf{21}, 656 (1965); R. Pudritz and
J. Silk, Ap. J.\textit{\ }\textbf{342}, 650 (1989); R. Kulsrud, R. Cen, J.
P. Ostriker, and D. Ryu, 1996, astro-ph/960714. E. R. Harrison, Mon. Not. R.
astron. Soc.\textit{\ }\textbf{147}, 279 (1970); $ibid$ Mon. Not. R. astron.
Soc. \textbf{165}, 185 (1973); A. D. Dolgov, Phys. Rev. D \textbf{48}, 2499
(1993); M. S. Turner and L. M. Widrow, Phys. Rev. D \textbf{30}, 2743
(1988); K. Enqvist and P. Olensen, Phys. Lett. \textbf{B329}, 195 (1994); T.
Vachaspati, Phys. Lett. \textbf{B265}, 258 (1991); A. D. Dolgov and J. Silk,
Phys. Rev. D \textbf{47}, 3144 (1993); C. Hogan, Phys. Rev. Lett\textit{. }%
\textbf{51}, 1488 (1983); J. Quashnock, A. Loeb and D. N. Spergel, Ap. J. 
\textbf{344}, L49 (1989); B. Ratra, Ap. J.\textit{\ }\textbf{391}, L1
(1992); $ibid$ Phys. Rev. D \textbf{45}, 1913. A. Kosowsky and A. Loeb,
astro-ph/9601055, Ap. J.

[6] V. N. Lukash and A. A. Starobinskii, Sov. Phys. JETP \textbf{39}, 742
(1974).

[7] W. Press, Ap. J. \textbf{205, }311 (1976).

[8] A. G. Doroshkevich, Y. B. Zeldovich and I. D. Novikov, Sov. Phys. JETP 
\textbf{26}, 408 (1968).

[9] C.W. Misner, Ap. J. \textbf{151}, 431 (1969); R.A. Matzner, Commun.
Math. Phys. \textbf{20}, 1 (1971).

[10] V. N. Lukash, Nuovo Cimento \textbf{B 35}, 269 (1976).

[11] L.P. Grishchuk, A.G. Doroshkevich and V.M. Yudin, Sov. Phys. JETP 
\textbf{69}, 1857 (1975).

[12] Y. Hosotani, Phys. Lett B \textbf{147}, 44 (1984); P.V. Moniz and J.
Mour M\~ao, Class. Quantum Grav. \textbf{8}, 1815 (1991); B.K. Darian and
H.P. K\"unzle, Class. Quantum Grav. \textbf{12}, 2651 (1995).

[13] M. B. Green, J. H. Schwarz and E. Witten, \textit{Superstring Theory, }%
Vol. I, Cambridge UP, Cambridge, (1987); E. S. Fradkin and A. A. Tseytlin,
Nucl. Phys. \textbf{B 261,} 1 (1985); C. G. Callan, E. J. Martinec and M. J.
Perry, Nucl. Phys. \textbf{B} \textbf{262}, 593 (1985); C. Lovelace, Nucl.
Phys.\textbf{\ B} \textbf{273}, 413 (1985); E. J. Copeland, A. Lahiri and D.
Wands, Phys. Rev. D \textbf{50}, 4868 (1994); ibid Phys. Rev. D \textbf{51,}
1569 (1995); N. A. Batakis and A. A. Kehagias, Nucl. Phys. B \textbf{449,}
248 (1995); N. A. Batakis, Phys. Lett. B \textbf{353,} 450 (1995); N. A.
Batakis, Nucl. Phys. B \textbf{353,} 39 (1995); J.D. Barrow and M.
Dabrowski, Phys. Rev. D \textbf{55}, 630 (1997).

[14] J.D. Barrow and K. Kunze, Phys. Rev. D \textbf{56},741 (1997).

[15] A. Vilenkin and P. Shellard, \textit{Cosmic Strings and Other
Topological Defects}, Cambridge UP, Cambridge, (1994).

[16] J. Stachel, Phys. Rev. D \textbf{21}, 2171 (1980).

[17] L. Marder, Proc. Roy. Soc. A \textbf{252}, 45; W. Israel, Phys. Rev. D 
\textbf{15,} 935 (1977).

[18] C.L. Bennett et al, Ap. J. Lett. \textbf{464,} L1 (1996).

[19] L. Landau and E.M. Lifshitz, \textit{The Classical Theory of Fields},
4th ed., Pergamon: Oxford (1975).

[20] Y. B. Zeldovich, Sov. Astron\textit{. }\textbf{13}, 608 (1970).

[21] I. S. Shikin, Sov. Phys. JETP \textbf{36}, 811 (1972).

[22] C. B. Collins, Comm. Math. Phys\textit{. }\textbf{27}, 37 (1972).

[23] J.D. Barrow, P.G. Ferreira, and J. Silk, Phys. Rev. Lett. \textbf{78},
3610 (1996)

[24] K.S. Thorne, Ap. J.\textit{\ }\textbf{148}, 51 (1967).

[25] A. G. Doroshkevich, Astrophysics \textbf{1}, 138 (1967).

[26] A. D. Linde and M.I. Zelnikov, Phys. Lett. B \textbf{215, }59 (1988).

[27] J. Yearsley and J.D. Barrow, Class. Quantum Grav. \textbf{13}, 2693
(1996).

[28] K. Jacobs, Ap. J.\textit{\ }\textbf{155}, 379 (1969); V.A. Ruban, Sov.
Astron. \textbf{26,} 632 (1983); ibid \textbf{29}, 5 (1985); V.G. LeBlanc,
D. Kerr and J. Wainwright, Class. Quantum Grav. \textbf{12}, 513 (1995).

[29] R. Maartens, G.F.R. Ellis and W. R. Stoeger, Phys. Rev. D \textbf{51},
1525 (1995).

[30] J. D. Barrow, Phys. Rev. D \textbf{51}, 3113 (1995).

[31] C. B. Collins and S. W. Hawking, Ap. J\textit{. }\textbf{181}, 317
(1972).

[32] A. G. Doroshkevich, V. N. Lukash, and I. D. Novikov, Sov. Phys. JETP%
\textit{\ }\textbf{37}, 739 (1973).

[33] J. D. Barrow, R. Juszkiewicz, and D. H. Sonoda, Mon. Not. R. Astron. Soc%
\textit{. }\textbf{213}, 917 (1985); J.D. Barrow, Can. J. Phys. \textbf{164, 
}152 (1986); E. F. Bunn, P. G. Ferreira, and J. Silk, Phys. Rev. Lett. 
\textbf{77}, 2883 (1996).

[34] J.D.Barrow and D. Sonoda, Phys. Rep\textit{.} \textbf{139}, 1 (1986).

[35] S.W. Hawking and G.F.R. Ellis, \textit{The Large Scale Structure of
Space-time}, Cambridge UP, Cambridge, 1972.

[36] M. White, D. Scott, and J. Silk, Annu. Rev. Astron. Astrophys. \textbf{%
32}, 319 (1994).

[37] L.F. Abbott and M. Wise, Nucl. Phys. B \textbf{244,} 541 (1984); J.D.
Barrow, A.B. Burd and D. Lancaster, Class. Quantum Grav. \textbf{3}, 551
(1985); F. Lucchin and S. Matarrese Phys. Rev. D \textbf{32}, 1316 (1985);
J.D. Barrow Phys. Lett. B (1987); ibid Quart. Jl. Roy. astron. Soc. \textbf{%
29}, 101 (1988).

[38] R. Wald, Phys. Rev. D \textbf{28}, 2118 (1983); J.D. Barrow, in \textit{%
The Very Early Universe, }eds. G. Gibbons, S.W. Hawking and S.T.C. Siklos,
Cambridge UP, Cambridge, 1983; J.D. Barrow, Phys. Lett.\textit{\ }B \textbf{%
180}, 335 (1987).

[39] J. D. Barrow, Mon. Not. R. astron. Soc\textit{. }\textbf{175}, 359
(1976).

[40] S.W. Hawking and R.J. Tayler, Nature \textbf{309}, 1278 (1966)

[41] E. Mart\'inez-Gonzalez and J.L. Sanz, Astron. Astrophys.\textbf{\ 300},
346 (1995).

[42] A. Kogut, G. Hinshaw and A. Banday, Phys. Rev. D\textbf{\ 55,} 1901
(1997).

[43] J. Adams, U.H. Danielsson, D. Grasso, and H. Rubinstein, Phys. Lett. B 
\textbf{388}, 253 (1996).

[44] A. Kosowsky and A. Loeb,  Astrophys. J. \textbf{469}, 1 (1996).

[45] J.D. Barrow and J. Levin, Chaos in the Einstein-Yang-Mills Equations,
gr-qc/9706065.

\end{document}